\documentclass[twocolumn,showpacs]{revtex4}
\usepackage{amsmath}
\usepackage{amssymb}
\usepackage{graphicx}
\usepackage{bm}
\usepackage{dcolumn}

\begin{document}

\title{Nature of superconducting state in the new phase in (TMTSF)$_{2}$PF$%
_{6}$ under pressure.}
\author{L. P. Gor'kov}
\author{P. D. Grigoriev}
\altaffiliation[Also at ]{L. D. Landau Institute for Theoretical Physics, Chernogolovka, Russia}
\email{pashag@itp.ac.ru}
\affiliation{National High Magnetic Field laboratory, Florida State University,
Tallahassee, Florida}
\date{\today }

\begin{abstract}
The unusual phase has been recently observed in the organic material (TMTSF)$%
_{2}$PF$_{6}$, where superconductivity (SC) coexists with spin-density wave
(SDW) in the pressure interval $p_{c1}<p<p_{c}$ below the first order
transition into SC or normal metal phase. Assuming that the coexistence
takes place on the microscopic scale, we consider the properties of the
intermediate phase. We show that the new superconducting state inside SDW
phase just above $p_{c1}$ must bear a triplet pairing.
\end{abstract}

\maketitle

Below the critical pressure, $p_{c}$, destroying spin-density wave (SDW) in
the quasi-one-dimensional (Q1D) organic compound (TMTSF)$_{2}$PF$_{6}$ a
pressure interval $p_{c1}<p<p_{c}$\ has been discovered,\cite%
{LeeTripletMany,Vuletic} in which the dielectric SDW and metallic/SC regions
coexist spatially. The details of this coexistence are not entirely
determined experimentally; in particular, the domain sizes of the coexisting
phases remain unknown. While in \cite{Vuletic} a macroscopic size was
assumed for domains, the reason for macroscopic coexistence at fixed
pressure is unclear yet. Instead of this, the spatially inhomogeneous phase,
called the soliton phase (SP), has been assumed\cite{GG} in the pressure
region $p_{c1}<p<p_{c}$. The emergent SP is then ascribed to the appearance
of metallic domain walls above $p_{c1}$. This phenomenon has been first
proposed for the charge-density waves.\cite{BGS} The experimental data on NMR%
\cite{Brown} and on AMRO\cite{LeeBrown} about the domain size do not
contradict the assumption of Ref. [\onlinecite{GG}].

One of the most interesting questions in this context is the
question about the origin and properties of the SC in this new
state. As it was shown recently,\cite{Orsay} superconductivity
appears first at $p_{c1}$; at higher pressure $T_{c}^{SC}$ increases
and reaches the value of SC transition temperature in the metallic
state. The mechanism and the type of SC in the normal phase (above
the critical pressure, $p_{c},$ for the first order phase
transition) still remains unknown, though some arguments in favor of
the triplet pairing have been suggested.\cite{GorkovJerome}\ In the
new
intermediate state ($p_{c1}<p<p_{c}$) the absence of the Knight shift change%
\cite{LeeKnightShift} and too high critical field $H_{c2}$ compared to the
values of critical temperature\cite{LeeTripletMany,LeeAngularHc} attract
special attention. In the present letter we address the issue of the type of
SC pairing in the intermediate phase.

Although in Ref. [\onlinecite{GG}] the onset of SP was suggested at $%
p>p_{c1} $, the alternative destructive mechanism of the gapped SDW state
could be realized as a gradual formation of electron-hole ungapped pockets
when pressure enhances the "antinesting" term of the quasi-1D electronic
spectrum in (TMTSF)$_{2}$PF$_{6}$ (for CDW such a mechanism was discussed in
\cite{GL,BGL}). It turns out that close to $p_{c1}$: $p-p_{c1}\ll p_{c}$,
the SC onset can be studied analytically for the two scenarios: of weakly
overlapping solitons in SP or at the appearance of small ungapped e-h
pockets on the background of the homogeneous SDW. The main result below is
that close to $p_{c1}$ in both scenarios the low-temperature Cooper
instability exists only for the triplet pairing.

The quasi-1D compound (TMTSF)$_{2}$PF$_{6}$ in normal state is characterized
by the two open Fermi surface (FS) sheets with the spectrum
\begin{equation}
\varepsilon (\mathbf{k})=v_{F}(\left\vert k_{x}\right\vert -k_{F})+t_{\perp
}(\mathbf{k}_{\perp }).  \label{1}
\end{equation}%
In the SC state the Gor'kov order parameter at each (left or right) FS has
the form%
\begin{equation}
\begin{split}
f_{\alpha \beta }^{LR}(\mathbf{r})& =<\hat{\Psi}_{\alpha }^{L}(\mathbf{r})%
\hat{\Psi}_{\beta }^{R}(\mathbf{r})>; \\
f_{\alpha \beta }^{RL}(\mathbf{r})& =<\hat{\Psi}_{\alpha }^{R}(\mathbf{r})%
\hat{\Psi}_{\beta }^{L}(\mathbf{r})>.
\end{split}
\label{GF1}
\end{equation}%
The spatial inversion symmetry in (TMTSF)$_{2}$PF$_{6}$ allows to classify
the pairing type by the symmetry of the order parameters in Eq. (\ref{GF1})
: $f_{\alpha \beta }^{LR}=\pm f_{\alpha \beta }^{RL}$, where the sign ($\pm $%
) depends on whether the SC pairing has singlet (+) or triplet (-)
character. For simplicity, we use the mean-field model, in which
only the backward scattering matrix element, $g_{1}$, between
electrons on the opposite sheets contributes to the SC
paring.\cite{Comm1} The convenience of such a model is that in the
metallic phase the Cooper instability would always manifest itself
at some $T_{c}^{SC}$ for triplet or singlet pairing depending on the
sign of the coupling constant $g_{1}$ in the familiar
relation\cite{Suhl,AGD}%
\begin{equation}
1=\left[ g_{1}\ln (\overline{\omega }/T_{c}^{SC})\right] ^{2},  \label{Tc1}
\end{equation}%
where $\overline{\omega }$ is a proper cutoff, and $g_{1}$ is the matrix
element of the backward scattering interaction multiplied by the density of
states at the Fermi level.

Before to apply the Cooper instability analysis to the phase with the SDW,
one needs first to determine the wave functions and the energy spectrum of
the latter. To achieve this goal we generalized the approach developed for
CDW\cite{BGL} to the SDW case. In particular, this approach allows the
treatment of the homogeneous SDW and of the SP on equal footing. As shown in
Ref. \cite{BGL}, the spectrum of Eq. (\ref{1}) allows the exact mapping of
the anisotropic Q1D problem onto the purely 1D one, where sophisticated
methods for studying solitons have been developed.\cite{BrazKirovaReview} As
in Ref. [\onlinecite{BGL}], we consider the general case of SDW order
parameter, $\hat{\Delta}_{SDW}(\mathbf{r})$, acquiring spatial modulation in
the presence of the soliton walls:
\begin{equation}
\hat{\Delta}_{SDW}(\mathbf{r})=\Delta _{SDW}(x)\cos \left( \mathbf{Q}\mathbf{%
r}\right) (\vec{\hat{\sigma}}\mathbf{l}).  \label{OrdeParameter}
\end{equation}%
The Shroedinger equation writes $\hat{H}_{\mathbf{k}}^{0}\boldsymbol{\Psi }_{%
\mathbf{k}}=\varepsilon _{\mathbf{k}}\boldsymbol{\Psi }_{\mathbf{k}}$ with
the Hamiltonian
\begin{equation}
\hat{H}_{\mathbf{k}}^{0}=%
\begin{pmatrix}
\hat{\varepsilon}(\mathbf{k}_{\perp })-\frac{iv_{F}d}{dx}; & \Delta
_{SDW}(x)(\vec{\hat{\sigma}}\mathbf{l}) \\
\Delta _{SDW}^{\ast }(x)(\vec{\hat{\sigma}}\mathbf{l}); & \hat{\varepsilon}(%
\mathbf{\mathbf{k}_{\perp }-Q_{\perp }})+\frac{iv_{F}d}{dx}%
\end{pmatrix}
\label{Ham1}
\end{equation}%
and with the four-component (spin) wave function
\begin{equation}
\boldsymbol{\Psi }_{\mathbf{k}}\equiv \binom{\boldsymbol{\psi }_{\mathbf{k}%
}^{R}(x)}{\boldsymbol{\psi }_{\mathbf{k-Q}}^{L}(x)},~\boldsymbol{\psi }_{%
\mathbf{k}}^{R(L)}(x)=\binom{{\psi }_{\mathbf{k\uparrow }}^{R(L)}(x)}{{\psi }%
_{\mathbf{k\downarrow }}^{R(L)}(x)},  \label{WaveFunction1}
\end{equation}%
which combines the electron wave functions on the right and left Fermi
surface sheets, denoted by $R\left( L\right) $ superscripts. Transformation%
\begin{eqnarray}
{\psi }_{\mathbf{k}_{\perp }\alpha }^{R}(x) &=&\exp \left\{ ix\left[
k_{x}-\varepsilon _{-}(\mathbf{k}_{\perp })/v_{F}\right] \right\} {\psi }%
_{\alpha }^{R}(x)  \label{Transform} \\
{\psi }_{\mathbf{k}_{\perp }-\mathbf{Q}_{\perp }\alpha }^{L}(x) &=&\exp
\left\{ -ix\left[ k_{x}-\varepsilon _{-}(\mathbf{k}_{\perp })/v_{F}\right]
\right\} {\psi }_{\alpha }^{L}(x),  \notag
\end{eqnarray}%
where
\begin{equation}
\varepsilon _{\pm }(\mathbf{k}_{\perp })=\left[ t(\mathbf{k}_{\perp })\pm t({%
\mathbf{k}_{\perp }-\mathbf{Q}_{\perp }})\right] /2,  \label{ts}
\end{equation}%
reduces the Hamiltonian (\ref{Ham1}) to
\begin{equation}
\hat{H}_{1D}=%
\begin{pmatrix}
-iv_{F}d/dx & \Delta (x)(\vec{\hat{\sigma}}\mathbf{l}) \\
\Delta ^{\ast }(x)(\vec{\hat{\sigma}}\mathbf{l}) & iv_{F}d/dx%
\end{pmatrix}%
.  \label{Ham1D}
\end{equation}%
The eigenvalues of the 3D problem (\ref{Ham1}) are
\begin{equation}
\varepsilon _{\lambda ,\mathbf{k}_{\perp }}=E_{\lambda }+\varepsilon _{+}(%
\mathbf{k}_{\perp }),  \label{eps3D}
\end{equation}%
where the index $\lambda $\ numerates the eigenvalues of the 1D Hamiltonian (%
\ref{Ham1D}) for a periodic soliton lattice $\Delta _{SDW}(x)$. Finding $%
\Delta _{SDW}(x)$\ is a separate problem, that can be solved exactly for the
commensurate case\cite{BGK} and in the limit of a single soliton.\cite{Braz}
For homogeneous SDW $\Delta _{SDW}(x)=const$\ the analysis of Eqs. (\ref%
{Ham1}),(\ref{WaveFunction1}) can be easily performed in the momentum
representation. The quasiparticle energy spectrum (\ref{eps3D}) than becomes
\begin{equation}
\varepsilon _{1,2}\left( {\boldsymbol{k}}\right) \equiv \varepsilon ^{+}({%
\boldsymbol{k}}_{\perp })\pm \sqrt{\xi ^{2}+|\Delta _{{\boldsymbol{SDW}}%
}|^{2}},  \label{epsM}
\end{equation}%
where $\xi \equiv v_{F}\left( \left\vert k_{x}\right\vert -k_{F}\right)
-\varepsilon ^{-}({\boldsymbol{k}}_{\perp })$.

The idea behind the calculation in both cases is that at $p>p_{c1}$ a branch
of the energy spectrum crosses the chemical potential. For a network of the
rarefied soliton walls, a single soliton wall may be treated as metallic
sheets\cite{BGS,GG} with the thickness $d\sim \xi _{0}=\hbar v_{F}/T_{SDW}$.
At higher pressure solitons will overlap and form a 3D metallic band that
lies inside the SDW gap. For the pockets' scenario at $p>p_{c1}$, the
transverse dispersion $\varepsilon _{+}(\mathbf{k}_{\perp })$ in Eq. (\ref%
{eps3D}) becomes greater than the SDW energy gap, forming first open
electron-hole pockets of the form
\begin{equation}
\varepsilon (\mathbf{k})=\pm \delta \pm \left[ a_{1}\left( \Delta \mathbf{k}%
_{\perp }\right) ^{2}+b_{1}\xi ^{2}\right] ,  \label{pocket1}
\end{equation}%
where%
\begin{eqnarray*}
\delta  &\equiv &\left\vert \Delta _{SDW}-t{_{\perp }}\left( {\mathbf{k}}%
_{0}\right) \right\vert \ll \Delta _{SDW}, \\
a_{1} &\sim &t{_{\perp }^{\prime }b}^{2}\text{ and }b_{1}\approx 1/2\Delta
_{SDW}.
\end{eqnarray*}%
In each case the formed small "Fermi surface" is subject to the examination
for a possible Cooper instability. Such an instability, should it occurs at
some low temperature, would signify the possibility for onset of SC pairing.

\begin{figure}[tbh]
\includegraphics[width=0.49\textwidth]{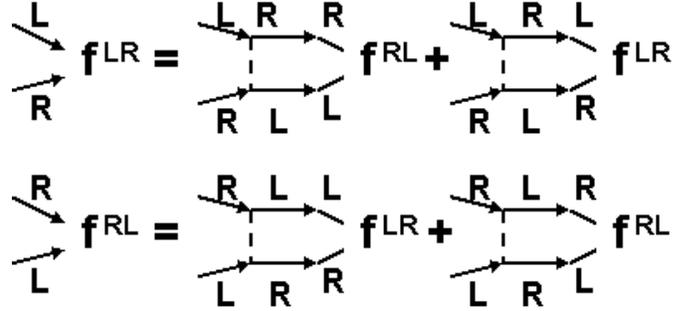}
\caption{The diagram equations for the functions $f_{\protect\alpha \protect%
\beta }^{LR}$ and $f_{\protect\alpha \protect\beta }^{RL}$ in the presence
of SDW. The solid lines represent the electron Green functions: $G_{i\protect%
\omega _{n}}^{R(L)R(L)}$. The dash lines represent the short-range
interaction (in our case the backward scattering) of electrons.}
\label{Fig1Eq2}
\end{figure}

Analysis of the Cooper ladder diagrams with the interaction $g_{1}$ in Eq. (%
\ref{Tc1}) can be carried out using the standard methods for
searching logarithmically divergent terms at $T\rightarrow
0$.\cite{AGD} In our case calculations are more tedious being
complicated by the underlying SDW structure. We briefly sketch the
main steps in the calculations. The corresponding diagram equations
are shown in Fig. \ref{Fig1Eq2}. The lines in the Cooper bubble
stand for the proper Greens functions. Compared to pairing in the
normal state, there are additional terms that come from
"non-diagonal" Greens functions, $G^{RL}$ and $G^{LR}$, when two
Fermi surface sheets of Eq. (\ref{1}) mix together in the SDW(CDW)
presence. The $4\times 4$ (spin) matrix Greens functions can be
presented in the general form
\begin{equation}
\hat{G}_{i\omega _{n},}(\mathbf{r},\mathbf{r}^{\prime })=-\sum_{\lambda ,%
\mathbf{k}_{\perp }}\frac{\boldsymbol{\Psi }_{\mathbf{k}}^{\dagger }(\mathbf{%
r}^{\prime })\otimes \boldsymbol{\Psi }_{\mathbf{k}}(\mathbf{r})}{i\omega
_{n}-\varepsilon _{\lambda ,\mathbf{k}_{\perp }}},  \label{SolG}
\end{equation}%
where $\boldsymbol{\Psi }_{\mathbf{k}}(\mathbf{r})$ is given by Eq. (\ref%
{WaveFunction1}).

Calculations are transparent for a homogeneous SDW(CDW). We discuss the case
in which small e-h pockets (\ref{pocket1}) get formed at some point $\mathbf{%
k}_{0}$ on the FS. In the momentum space the Green functions entering the
Cooper block in Fig. 1 write down as

\begin{equation}
g^{RR(LL)}({\boldsymbol{k}},\omega )=\frac{i\omega -\varepsilon ({%
\boldsymbol{k}})}{[i\omega -\varepsilon _{1}({\boldsymbol{k}})][i\omega
-\varepsilon _{2}({\boldsymbol{k}}]}  \label{gRR}
\end{equation}%
where $\omega \equiv \omega _{n}=\left( 2n+1\right) \pi T$, $\varepsilon ({%
\boldsymbol{k}})$ is given by Eq. (\ref{1}) and $\varepsilon _{1,2}$ are
given by Eq. (\ref{epsM}). The nondiagonal Green functions $\hat{g}^{LR(RL)}(%
{\boldsymbol{k}},\omega )=(\vec{\hat{\sigma}}\vec{l})g^{LR(RL)}({\boldsymbol{%
k}},\omega )$, with
\begin{equation}
g^{LR}({\boldsymbol{k}},\omega )=\frac{\Delta _{SDW}}{[i\omega -\varepsilon
_{1}({\boldsymbol{k}})][i\omega -\varepsilon _{2}({\boldsymbol{k}}]}
\label{gLR}
\end{equation}%
and $g^{RL}({\boldsymbol{k}},\omega )=\left[ g^{LR}({\boldsymbol{k}},-\omega
)\right] ^{\ast }$.

The sum of two Cooper bubbles in the right-hand part of equations, shown
schematically in Fig. 1, writes down as
\begin{eqnarray}
&&\hat{f}^{LR}=-Tg_{1}\sum_{\mathbf{k},\omega }\left[ g^{RR}({\boldsymbol{k}}%
,\omega )\hat{f}^{RL}g^{LL}(-{\boldsymbol{k}},-\omega )\right.
\label{Bubble} \\
&&+\left. (\vec{\hat{\sigma}}\vec{l})g^{LR}({\boldsymbol{k}},\omega )\hat{f}%
^{LR}(\vec{\hat{\sigma}}\vec{l})^{T}g^{RL}(-{\boldsymbol{k}},-\omega )\right]
.  \notag
\end{eqnarray}%
The ultraviolet logarithmic divergence in (\ref{Bubble}) comes only from the
first line; it would give the term $\left\vert g_{1}\right\vert \ln \left(
\bar{\omega}/C\Delta _{SDW}\right) $, where $C=const\sim 1$. Once electrons
and hole pockets open at the Fermi level, they lead to the appearance of the
low-energy divergence that contribute an additional logarithmical divergent
term $\sim \ln \left( \sqrt{\delta \Delta _{SDW}}/T\right) $, where $\delta $
determines the size of small pockets. The total equation (\ref{Bubble}) for
the SC transition temperature than rewrites as%
\begin{equation}
\ln \left( C\Delta _{SDW}/T_{SC}\right) =A\ln \left( \sqrt{\delta \Delta
_{SDW}}/T\right) .  \label{Log}
\end{equation}%
(Remember that $\Delta _{SDW}\approx 12K\gg T_{SC}\approx 1K$.) Calculations
of this contribution from the single pocket go as follows. Consider Eq. (\ref%
{pocket1}) for one pocket. The energy level $\varepsilon _{1}(\mathbf{k})$
crosses the chemical potential at the point $\mathbf{k}=\mathbf{k}_{0}$ to
form a small FS at pressure slightly above $p_{c1}$. Then $\varepsilon _{2}(%
\mathbf{k})\approx 2\Delta _{SDW}$ at $\mathbf{k}$ near $\mathbf{k}_{0}$.
Similarly, taking in the nominator of Greens functions $\varepsilon (\mathbf{%
k})\approx \Delta _{SDW}$ at $\mathbf{k}$ near $\mathbf{k}_{0}$ and
substituting the simplified Greens functions (\ref{gRR}),(\ref{gLR})
into (\ref{Bubble}) we find a familiar form of the logarithmic
divergence at low temperature coming from the poles of the Greens
functions:
\begin{equation*}
\sum_{\mathbf{k},\omega }\frac{T}{\omega ^{2}+\varepsilon _{1}^{2}({%
\boldsymbol{k}})}=\sum_{\mathbf{k}}\frac{\tanh \left[ \varepsilon _{1}({%
\boldsymbol{k}})/2T\right] }{2\varepsilon _{1}({\boldsymbol{k}})}.
\end{equation*}%
Substituting (\ref{pocket1}) for $\varepsilon _{1}({\boldsymbol{k}})$ and
changing the variables $a\left( \Delta \mathbf{k}_{\perp }\right)
^{2}\rightarrow y^{2}$ and $b\left( k_{x}-k_{F}\right) ^{2}\rightarrow x^{2}$
we obtain the integral of the form%
\begin{eqnarray*}
&&\frac{1}{\sqrt{a_{1}b_{1}}}\int \frac{\tanh \left[ \varepsilon _{1}({%
\boldsymbol{x,y}})/2T\right] }{2\varepsilon _{1}({\boldsymbol{x,y}})}\frac{%
dxdy}{\left( 2\pi \right) ^{2}} \\
&=&\frac{1}{\sqrt{a_{1}b_{1}}}\int_{0}^{\delta }\frac{\tanh \left[ \left(
\delta -r^{2}\right) /2T\right] }{\delta -r^{2}}\frac{dr^{2}}{8\pi } \\
&&+\frac{1}{\sqrt{a_{1}b_{1}}}\int_{0}^{\Delta _{SDW}}\frac{\tanh \left[
\left( \delta -r^{2}\right) /2T\right] }{\delta -r^{2}}\frac{dr^{2}}{8\pi }
\\
&\sim &\frac{\ln \left[ \sqrt{\Delta _{SDW}\delta }/T\right] }{4\pi }.
\end{eqnarray*}

Returning to Eq. (\ref{Log}), the value of the prefactor $A$, which
defines the SC transition temperature, is just a number. Most
remarkable, however, is the observation that $A$ drastically depends
on the type of pairing. For
spin-singlet paring the spin structure of the SC order parameter $\hat{f}%
^{LR}=\hat{f}^{LR}=i\hat{\sigma}_{y}f^{LR}$, and using $\hat{\sigma}_{y}(%
\vec{\hat{\sigma}}\vec{l})^{T}=-$ $(\vec{\hat{\sigma}}\vec{l})\hat{\sigma}%
_{y}$, one rewrites equation (\ref{Bubble}) as%
\begin{eqnarray}
1 &=&-Tg\sum_{\mathbf{k},\omega }\left[ g^{RR}({\boldsymbol{k}},\omega
)g^{LL}(-{\boldsymbol{k}},-\omega )\right.   \label{Bs} \\
&&-\left. g^{LR}({\boldsymbol{k}},\omega )g^{RL}(-{\boldsymbol{k}},-\omega )%
\right] .  \notag
\end{eqnarray}%
The second line in this equation acquires the sign "$-$" due to the spin
structure of the background SDW phase, which is in contrast to the SC on the
CDW background. This difference in the sign leads to the cancelation in the
main approximation of the low-energy logarithmic singularity in (\ref{Bs})
for the chosen pocket at $\mathbf{k}=\mathbf{k}_{0}$. This results in a
smallness of the factor $A\sim \delta /\Delta _{SDW}$ before the logarithm
in the r.h.s. of Eq. (\ref{Bubble}).

This cancelation may not occur for triplet pairing. Substituting the spin
structure of triplet order parameter, $\hat{f}^{LR}=\left( \mathbf{\hat{%
\sigma}\vec{d}}\right) \hat{\sigma}_{y}f^{LR}$, together with $f^{RL}=-f^{LR}
$\ into (\ref{Bubble}) and using $(\vec{\hat{\sigma}}\vec{l})\left( \mathbf{%
\hat{\sigma}\vec{d}}\right) \hat{\sigma}_{y}(\vec{\hat{\sigma}}\vec{l}%
)^{T}=\left( \mathbf{\hat{\sigma}\vec{d}}\right) \hat{\sigma}_{y}-2\left(
\mathbf{\vec{d}}\vec{l}\right) (\vec{\hat{\sigma}}\vec{l})\hat{\sigma}_{y}$
we obtain in the right hand part of Eq. (\ref{Bubble})%
\begin{eqnarray}
&&T\sum_{\mathbf{k},\omega }\left[ -g^{RR}({\boldsymbol{k}},\omega )\left(
\mathbf{\hat{\sigma}\vec{d}}\right) g^{LL}(-{\boldsymbol{k}},-\omega
)\right.   \label{Bt} \\
&&\left. +g^{LR}({\boldsymbol{k}},\omega )g^{RL}(-{\boldsymbol{k}},-\omega
)\left\{ \left( \mathbf{\hat{\sigma}\vec{d}}\right) -2\left( \mathbf{\vec{d}}%
\vec{l}\right) (\vec{\hat{\sigma}}\vec{l})\right\} \right] .  \notag
\end{eqnarray}%
We see that the main infrared divergent terms cancel each other only if $%
\mathbf{\vec{d}\perp }\vec{l}$. For $\mathbf{\vec{d}\parallel }\vec{l}$ the
factor $A$ is the same, as in the case of the CDW background.

Analysis for the onset of SC in the soliton wall scenario goes through in
the similar fashion. The logarithmic singularity of the Cooper type via the
isolated soliton wall sheets has already been discussed for CDW.\cite{GL}
For SDW one has to return to Eq. (\ref{Bubble}) and the wave functions (\ref%
{WaveFunction1}) making use of the exact single soliton solution\cite{Braz}.
One can easily check that similar cancelation in the nominator depending on
the spin structure happens in this scenario also.

To summarize, we have shown that at either way the SDW is being
destroyed by pressure above $p_{c1}$, SC in this new state is
expected to bear triplet character. This result also shows the
remarkable difference between SDW and CDW coexisting with
superconductivity on a single conducting band. Our results, although
have been derived assuming $\left\vert p-p_{c1}\right\vert \ll
p_{c1}$, should extend over a considerable part of the new phase in
(TMTSF)$_{2}$PF$_{6}$ at $p_{c1}<p<p_{c}$ if there is no additional
phase transition at $p<p_{c}$.

The work was supported by NHMFL through the NSF Cooperative agreement No.
DMR-0084173 and the State of Florida, and (PG) in part, by DOE Grant
DE-FG03-03NA00066 and RFBR N 06-02-16551.


\begin{thebibliography}{99}
\bibitem{Vuletic} T. Vuletic, P. Auban-Senzier, C. Pasquier et al., Eur.
Phys. J. B \textbf{25}, 319 (2002).

\bibitem{LeeTripletMany} I. J. Lee, M. J. Naughton, G. M. Danner, and P. M.
Chaikin, Phys. Rev. Lett. \textbf{78}, 3555 (1997); I. J. Lee, P. M.
Chaikin, and M. J. Naughton, Phys. Rev. B \textbf{62}, R14 669 (2000); I. J.
Lee, P. M. Chaikin, and M. J. Naughton, Phys. Rev. Lett. \textbf{88}, 207002
(2002).

\bibitem{GG} L.P. Gor'kov, P.D. Grigorev, Europhys. Lett. \textbf{71}, 425
(2005).

\bibitem{BGS} S.A. Brazovskii, L.P. Gor'kov, J.R. Schrieffer, Physica
Scripta \textbf{25}, 423 (1982).

\bibitem{Brown} S.E. Brown, 2005 (unpublished).

\bibitem{LeeBrown} I. J. Lee, S. E. Brown, W. Yu, M. J. Naughton, and P. M.
Chaikin, Phys. Rev. Lett. \textbf{94}, 197001 (2005).

\bibitem{Orsay} The Orsay group, private communication, 2006 (unpublished).

\bibitem{GorkovJerome} L. P. Gor'kov and D. Jerome, J. Phys. (Paris) Lett.
\textbf{46}, L-643 (1985).

\bibitem{LeeKnightShift} I. J. Lee, S. E. Brown, W. G. Clark, M. J. Strouse,
M. J. Naughton, W. Kang, and P. M. Chaikin, Phys. Rev. Lett. \textbf{88},
017004 (2002); I.J. Lee, D. S. Chow, W. G. Clark, M. J. Strouse, M. J.
Naughton, P. M. Chaikin, and S. E. Brown, Phys. Rev. B \textbf{68}, 092510
(2003).

\bibitem{LeeAngularHc} I. J. Lee, P. M. Chaikin, and M. J. Naughton, Phys.
Rev. B \textbf{65}, 180502(R) (2002).

\bibitem{BGL} S.A. Brazovskii, L.P. Gor'kov, A.G. Lebed', Sov. Phys. JETP
\textbf{56}, 683 (1982) [Zh. Eksp. Teor. Fiz. \textbf{83}, 1198 (1982)].

\bibitem{GL} L.P. Gor'kov, A.G. Lebed', Journal de Physique \textbf{44},
C3-1531 (1983).

\bibitem{Comm1} We have checked that our result does not change in the
general case, when the forward scattering is also taken into
account.

\bibitem{Suhl} H. Suhl, B.T. Matthias, and L.R. Walker, Phys. Rev. Lett.
\textbf{3}, 552 (1959)

\bibitem{AGD} A.A. Abrikosov, L.P. Gor'kov and I.E. Dzyaloshinskii,
\textquotedblright Methods of quantum field theory in statistical
physics\textquotedblright , Dover Publications, INC., New York 1977.

\bibitem{BrazKirovaReview} S.A. Brazovskii and N.N. Kirova, Sov. Sci. Rev. A
Phys., \textbf{5}, 99 (1984).

\bibitem{BGK} S.A. Brazovskii, S.A. Gordyunin and N.N. Kirova, JETP Lett.
\textbf{31}, 451 (1980)[Pis'ma v ZhETF \textbf{31}, 486 (1980)].

\bibitem{Braz} S.A. Brazovskij, Sov. Phys. JETP \textbf{51}, 342 (1980).
\end{thebibliography}
\end{document}